\documentclass[aps,nofootinbib,prd,preprint,tightenlines,amsmath,amssymb]{revtex4-2}

\usepackage{bm}
\usepackage[colorlinks,anchorcolor=black,citecolor=violet,linkcolor=blue]{hyperref}
\usepackage[mathscr]{euscript}
\usepackage{graphicx}
\usepackage{multirow,slashed}
\usepackage{xcolor}

\begin{document}
\baselineskip=17pt \parskip=5pt

\preprint{NCTS-PH/2013}

\title{Exploring charm decays with missing energy in leptoquark models}

\author{Gaber Faisel$,^{1}$ Jhih-Ying Su$,^{2}$ and Jusak Tandean$^{3,4}$ \bigskip \\ \it
$^1$Department of Physics, Faculty of Arts and Sciences,
S\"uleyman Demirel University, Isparta 32260, Turkey \medskip \\
$^2$Department of Physics, University of Massachusetts Amherst, Massachusetts 01003, USA \medskip \\
$^3$Department of Physics, National Taiwan University, Taipei 106, Taiwan \medskip \\
$^4$Physics Division, National Center for Theoretical Sciences, Hsinchu 300, Taiwan \bigskip \\
\large\rm Abstract \medskip \\
\begin{minipage}{\textwidth} \baselineskip=17pt \parindent=3ex \small
We investigate the possibility that scalar leptoquarks generate consequential effects on
the flavor-changing neutral-current decays of charmed hadrons into final states with missing
energy ($\slashed E$) carried away by either standard model or sterile neutrinos.
We focus on scenarios involving the $R_2$, $\tilde R_2$, and $\bar S_1$ leptoquarks and take
into account various pertinent constraints, learning that meson-mixing ones and those
inferred from collider searches can be of significance.
We find in particular that the branching fractions of charmed meson decays $D\to M\slashed E$,
$M=\pi,\rho$, and $D_s\to K^{(*)}\slashed E$ and singly charmed baryon decays
$\Lambda_c^+\to p\slashed E$ and $\Xi_c\to\Sigma\slashed E$ are presently allowed to attain
the $10^{-7}$-$10^{-6}$ levels if induced by $R_2$ and that the impact of $\tilde R_2$ is
comparatively much less.
In contrast, the contributions of $\bar S_1$, which couples to right-handed up-type quarks and
the sterile neutrinos, could lead to branching fractions as high as order $10^{-3}$.
This suggests that these charmed hadron decays might be within reach of the BESIII and Belle II 
experiments or future super charm-tau factories and could serve as potentially
promising probes of leptoquark interactions with sterile neutrinos.
\end{minipage}}


\maketitle

\section{Introduction\label{intro}}

The flavor-changing neutral-current (FCNC) decays of charmed hadrons into a lighter hadron plus
missing energy ($\slashed E$) have long been anticipated in the literature~\cite{Burdman:2001tf,
Chen:2007cn,Chen:2007yn,Badin:2010uh,Mahmood:2014wna,deBoer:2015boa,Dorsner:2016wpm,Li:2018hgu,
MartinCamalich:2020dfe,Gabrielli:2016cut,Fabbrichesi:2020wbt,Su:2020yze,Bause:2020auq,
Bause:2020xzj} to be among likely environments in which to discover hints of new
physics (NP) beyond the standard model (SM).
In the SM these processes arise at short distance from the quark transition \,$c\to u\nu\bar\nu$,\,
which emits unobserved neutrinos ($\nu\bar\nu$) and is greatly suppressed because it proceeds from
loop diagrams and is subject to very efficient Glashow-Iliopoulos-Maiani
cancellation~\cite{Burdman:2001tf}.
The effects of long-distance physics on these decays have also been estimated to be
tiny~\cite{Burdman:2001tf}.
Beyond the SM there could be extra ingredients causing modifications to the SM component
and/or yield additional channels with one or more invisible nonstandard particles, which might
translate into substantially amplified rates detectable by upcoming quests.

Experimentally, there still has not been a lot of activity to look for charmed hadron
decays of this kind~\cite{Zyla:2020zbs}.
At the moment the sole result available is a limit on the branching fraction of charmed meson
decay \,$D^0\to\slashed E$,\, which has been set by the Belle Collaboration~\cite{Lai:2016uvj}.
Due to the importance of these processes as valuable tools in the quest for NP,
it is hoped that dedicated efforts will increasingly be made to pursue them.
Since a clean environment and sizable luminosity are crucial for such endeavors, it is then timely
that there are now heavy-flavor factories which are running and expectedly well-suited for them,
namely BESIII~\cite{Asner:2008nq} and Belle II~\cite{Kou:2018nap}.
In the future, further measurements with improved sensitivity would presumably be feasible, such as
at the proposed super charm-tau factories~\cite{Luo:2018njj,Barnyakov:2020vob} and Circular
Electron Positron Collider (CEPC) operated as a $Z$-boson factory~\cite{CEPCStudyGroup:2018ghi}.

The foregoing motivates us in this paper to explore these FCNC charm transitions with
missing energy in the contexts of relatively simple NP scenarios.
We entertain specifically the possibility that heavy leptoquarks (LQs) with spin 0 mediate
the NP contributions to the FCNCs.
Over the last several years LQs have attracted a good deal of attention because suggested models
containing them are among those that could offer the preferred explanations for the so-called
\mbox{$B$-physics} anomalies~\cite{Dorsner:2016wpm,Bifani:2018zmi}.
While more data are awaited in order to clarify whether or not these anomalies are attributable
to NP, it is therefore of interest to investigate if LQs can give rise to appreciable 
manifestations in the charm sector too.
There have been various analyses in the past, such as Refs.\,\,\cite{deBoer:2015boa,Dorsner:2016wpm,
Bause:2020xzj,Davies:1990sc,Leurer:1993em,Davidson:1993qk,Golowich:2007ka,Fajfer:2015mia,
Sahoo:2017lzi,Benbrik:2008ik,Fajfer:2008tm,deBoer:2017que,Bause:2019vpr,Bharucha:2020eup},
looking into the effects of LQs on FCNC charm processes, but \,$c\to u\slashed E$\, was covered 
by only a few~\cite{deBoer:2015boa,Dorsner:2016wpm,Bause:2020xzj}.
The outcomes of our work would be complementary to those of the latter.

Besides the scalar LQs, we will incorporate light right-handed sterile neutrinos into the theory.
They are singlets under the SM gauge groups and in the presence of the LQs can have renormalizable
links to SM quarks.
Our inclusion of right-handed neutrinos is well motivated for two reasons.
First, their existence will be necessary if measurements in the future establish that neutrinos
are Dirac in nature.
Second, as will be demonstrated later on, relative to those with SM fermions alone
the LQ-mediated FCNCs that involve the sterile neutrinos and SM quarks might exert considerably
enhanced influence on the charm decays of concern, especially if the quarks are also right-handed.
This latter LQ feature has not been treated previously in the \,$c\to u\slashed E$
context~\cite{deBoer:2015boa,Dorsner:2016wpm,Bause:2020xzj}.

With respect to the predictions of the models to be studied shortly, it will be useful to have
some idea about the extent to which they might be accessible by the aforementioned experiments.
Here we address this question briefly, in light of the currently scant details obtainable from 
the literature.
For \,$D^0\to\pi^0\slashed E$,\, its branching fraction could be tested by BESIII down to
$10^{-4}$ or lower~\cite{Ablikim:2019hff}, after it gathers a data sample of 20\,fb$^{-1}$
at center-of-mass energy \,$\sqrt s\simeq3.77$\,GeV.\,
No corresponding information exists for Belle II as far as we can tell, but a rough estimate may be
made based on the projected prospects of these ongoing efforts to discover \,$D^0\to\slashed E$.\,
Thus, since it is expected to improve on the Belle bound
\,${\cal B}\big(D^0\to\slashed E\big)<9.4\times10^{-5}$ \cite{Lai:2016uvj} by
a factor of seven~\cite{Kou:2018nap} and BESIII could strengthen it further to $10^{-6}$ with its 
final charm dataset~\cite{Ablikim:2019hff}, the Belle II sensitivity to
${\cal B}\big(D^0\to\pi^0\slashed E\big)$ may reach only around
\,$(10^{-4}/10^{-6})(9.4\!\times\!10^{-5}/7)\sim10^{-3}$\, if the ratios of efficiencies of
these machines to reconstruct the two modes are alike. 
Nevertheless, it is hoped that Belle II will ultimately be able to exceed this naive expectation.
Later in the future, the super charm-tau factories~\cite{Luo:2018njj,Barnyakov:2020vob} are
planned to attain total luminosities 100 times that of~BESIII, and so they could have 
the capability to access \,${\cal B}\big(D^0\to\pi^0\slashed E\big)\sim10^{-5}$\, or better.
On the other hand, the CEPC running at the $Z$ pole~\cite{CEPCStudyGroup:2018ghi} is proposed to
produce a few times more charmed hadrons than Belle II and hence might be somewhat superior to
the latter for probing this decay.
Given that the $D^0$ amount collected at these different facilities~\cite{Ablikim:2019hff,
Kou:2018nap,CEPCStudyGroup:2018ghi} is bigger than those of $D_{(s)}^+$ and charmed baryons,
their sensitivity to the other FCNC charmed-hadron transitions we will look at would probably
be similar or less.

The organization of the remainder of the paper is as follows.
In Sec.\,\ref{lqc} we describe the interactions of the new particles, namely the scalar LQs and
the light sterile neutrinos, with the SM fermions.
Concentrating on three distinct scalar LQs, in Secs.\,\ref{R2}, \ref{tR2}, and \ref{bS1}
we examine three separate scenarios, in each of which merely one of the LQs is responsible
for \,$c\to u\slashed E$.\,
In these sections, before presenting our numerical results, we first deal with the pertinent
limitations on the Yukawa parameters and masses of the LQs.
We point out particularly that meson-mixing constraints in the charm and down-type quark sectors are
potentially relevant, but may be avoided in certain instances.
Furthermore, we take into account restrictions inferred from collider searches.
Subsequently, we evaluate a variety of FCNC decays of the lightest charmed pseudoscalar-mesons
and singly charmed baryons manifesting the LQ effects on \,$c\to u\slashed E$.\,
In Sec.\,\ref{concl} we draw our conclusions.
In the Appendix, we provide general expressions for the rates of the hadron decay
modes under consideration.
For the meson channels we show that, assuming the invisible particles to have negligible masses, 
the formulas for the LQ-generated branching fractions can be determined with the aid of data on 
the corresponding semileptonic modes and without relying on the form factors in the mesonic matrix
elements.

\section{Leptoquark couplings to fermions\label{lqc}}

Among LQs that can have renormalizable interactions with SM fermions without violating
the conservations of baryon and lepton numbers and the SM gauge symmetries, there are four which
possess spin 0 and can at tree level contribute to the quark transition \,$c\to u\slashed E$\,
where the missing energy is carried away by either SM or sterile
neutrinos~\cite{Dorsner:2016wpm}.\footnote{In the recent literature covering the impact of NP on
\,$c\to u\slashed E$,\, the missing energy could alternatively be carried away by a single particle
such as QCD axion~\cite{MartinCamalich:2020dfe} or massless dark
photon~\cite{Gabrielli:2016cut,Fabbrichesi:2020wbt,Su:2020yze}.}
In the nomenclature of Ref.\,\cite{Dorsner:2016wpm}, these scalar LQs, with their assignments
under the SM gauge groups SU(3$)_{\rm color}\times{\rm SU}(2)_L\times{\rm U}(1)_Y$, are denoted by
\,$R_2\,(3,2,7/6)$, \,$\tilde R_2\,(3,2,1/6)$,\, $\bar S_1\,\big(\bar 3,1,-2/3\big)$,\,
and \,$S_3\,\big(\bar 3,3,1/3\big)$.\,
Here we pay attention to the first three because we have found that the couplings of $S_3$ are
comparatively more restrained than those of the other three LQs.
In terms of their components,
\begin{align}
R_2^{} & \,= \left(\!\begin{array}{c} R_2^{5/3^{\vphantom{|}}} \medskip \\
R_2^{2/3} \end{array}\!\right) , &
\tilde R_2^{} & \,= \left(\!\begin{array}{c} \tilde R_2^{2/3^{\vphantom{|}}} \medskip \\
\tilde R_2^{-1/3} \end{array}\!\right) , &
\bar S_1^{} & \,=\, \bar S_1^{-2/3} \,, ~~~~~
\end{align}
where the superscripts refer to their electric charges.

As for the right-handed neutrinos, we assume that there are three of them ($\textsl{\texttt N}_1$,
$\textsl{\texttt N}_2$, and $\textsl{\texttt N}_3$) and that they are of Dirac nature.   
Moreover, we suppose that they have masses which may be unequal but are sufficiently small to be 
neglected in the charmed-hadron processes of concern.
In addition, we take $\textsl{\texttt N}_{1,2,3}$ to be long-lived enough that they do not decay 
inside detectors.

We express the Lagrangian for the renormalizable interactions of $R_2$, $\tilde R_2$, and
$\bar S_1$ with SM fermions plus $\textsl{\texttt N}_{1,2,3}$ as
\begin{align} \label{Llq}
{\cal L}_{\textsc{lq}}^{} & \,=\,
{\tt Y}{}_{2,jy}^{\textsc{rl}}\, \overline{u_j^{}} R_2^{\textsc t} i\tau_2^{~} P_L^{} l_y^{}
+ \tilde{\textsc y}_{2,jy}^{\textsc{lr}}\, \overline{q_j^{}} \tilde R_2^{} P_R^{}
\textsl{\texttt N}_y^{}
+ \bar{\textsc y}_{1,jy}^{\textsc{rr}}\, \overline{u_j^{\textsc c}} P_R^{}
\textsl{\texttt N}_y^{} \bar S_1^{}
\,+\, \rm H.c. \,,
\end{align}
where ${\tt Y}{}_{2,jy}^{\textsc{rl}}$, $\tilde{\textsc y}_{2,jy}^{\textsc{lr}}$, and
$\bar{\textsc y}_{1,jy\,}^{\textsc{rr}}$  are generally complex elements of the LQ Yukawa matrices,
summation over family indices \,$j,y=1,2,3$\, is implicit, $q_j$ $(l_y)$ and $u_j$ symbolize
a left-handed quark (lepton) doublet and right-handed up-type quark singlet, $\tau_2^{}$ is
the second Pauli matrix, $P_{L,R}=(1\mp\gamma_5^{})/2$,\, and the superscript \textsc c
indicates charge conjugation.
In Eq.\,(\ref{Llq}), we introduce only the minimal ingredients which serve our purposes pertaining
to the \,$c\to u\slashed E$\, transitions to be studied.
Now we entertain three distinct possibilities each involving one of the LQs, taken to be heavy,
with the couplings specified above.

\section{\boldmath$R_2$ model\label{R2}}

Expanding the $R_2$ portion of Eq.\,(\ref{Llq}), we have
\begin{align} \label{LR2}
{\cal L}_{R_2}^{} & \,=\, {\texttt Y}_{2,jy}^{\textsc{rl}}\, \overline{{\cal U}_j^{}} P_L^{}
\Big( \ell_y^{} R_2^{5/3} - \nu_{\ell_y}^{} R_2^{2/3} \Big) \,+\, \rm H.c. \,, ~~~~~
\end{align}
where \,${\cal U}_{1,2,3}=u,c,t$\, and \,$\ell_{1,2,3}=e,\mu,\tau$\, represent mass eigenstates.
Given that the ordinary neutrinos in the decays of interest have vanishing masses and are not
detected experimentally, we can work with the states $\nu_{\ell_y}$ associated with $\ell_y$.

From ${\cal L}_{R_2}$, one can derive effective \,$|\Delta C|=1$\, quark-lepton
operators which at low energies are expressible as
\begin{align} \label{ucff'-R2}
{\cal L}_{uc\texttt{ff}'}^{} & \,=\, -\sqrt2\, G_{\rm F}^{}\, \textsc k_{\ell_x\ell_y}^{}
\overline u\gamma_\beta^{} P_R^{}c\, \Big( \overline{\nu_{\ell_x}^{}} \gamma^\beta P_L^{}
\nu_{\ell_y}^{} + \overline{\ell_x^{}} \gamma^\beta P_L^{} \ell_y^{} \Big) \,+\, \rm H.c. \,, ~~~~~
\end{align}
where $G_{\rm F}$ is the Fermi constant, \,$x,y=1,2,3$\, are summed over, and
\begin{align} \label{Cll'}
\textsc k_{\ell_x\ell_y}^{} & \,=\, \frac{\hat v^2\, {\texttt Y}_{2,1y}^{\textsc{rl}}
{\texttt Y}_{2,2x}^{\textsc{rl}*}}{2 m_{R_2}^2} ~~~~ ~~~
\end{align}
is a dimensionless coefficient, with \,$\hat v=2^{-1/4}G{}_{\rm F}^{-1/2}\simeq246$\,GeV\, and
$m_{R_2}$ being the mass of $R_2$.
They induce FCNC charmed-hadron decays with missing energy as well as those with charged
leptons in the final state.
Before treating the former processes, we look at some potentially important constraints on
the LQ parameters in Eq.\,(\ref{Cll'}).

It is long known that scalar-LQ interactions could influence the mixing of charmed mesons $D^0$
and $\bar D^0$ via \,$|\Delta C|=2$\, four-quark operators arising from box diagrams
\cite{deBoer:2015boa,Dorsner:2016wpm,Davies:1990sc,Leurer:1993em,Davidson:1993qk,Golowich:2007ka,
Fajfer:2015mia,Sahoo:2017lzi,Bause:2019vpr}.
In the presence of ${\cal L}_{R_2}$ in Eq.\,(\ref{LR2}), the loops contain the SM charged
and neutral leptons, besides $R_2$.
This results in the effective Hamiltonian~\cite{Dorsner:2016wpm}
\begin{align} \label{Hucuc-R2}
{\cal H}_{|\Delta C|=2}^{R_2} & \,=\, \frac{\big( \mbox{$\sum$}_x^{} \texttt Y_{2,1x}^{\textsc{rl}}
\texttt Y_{2,2x}^{\textsc{rl}*} \big)\raisebox{1pt}{$\!^2$} }{64\pi^2\, m_{R_2}^2}\,
\overline u\gamma^\beta P_R^{}c\, \overline u\gamma_\beta^{}P_R^{}c \,+\, {\rm H.c.} ~~~~~
\end{align}
It affects the mixing observable
\,$\Delta m_D^{} = |\langle\bar D^0|{\cal H}_{|\Delta C|=2}|D^0\rangle|\tilde r/m_{D^0}$,\,
where \,$\tilde r=0.74$\, accounts for the renormalization-group running of the coefficient in
Eq.\,(\ref{Hucuc-R2}) from the scale \,$m_{R_2}=2$ TeV\, down to 3 GeV~\cite{Golowich:2007ka}.
Our choice for $m_{R_2}$ is consistent with the negative outcome of a recent direct search at
the LHC for scalar LQs decaying fully into a quark and an electron (muon), which has excluded
the mass region below 1.8\,(1.7)\,TeV~\cite{Aad:2020iuy}.

Employing
\,$\langle\bar D^0|\overline u\gamma^\kappa P_Rc\, \overline u\gamma_\kappa^{}P_Rc|D^0\rangle =
0.0805(57)\rm\,GeV^4$\,
at the scale of 3 GeV from a lattice QCD computation~\cite{Bazavov:2017weg} and the empirical value
\,$\Delta m_D^{\rm exp}=\big(95_{-44}^{+41}\big)\times10^8$/s~\cite{Zyla:2020zbs},
and assuming that the $R_2$ contribution saturates the latter, as the SM prediction suffers from
a sizable hadronic uncertainty~\cite{Golowich:2007ka}, we then get the 2$\sigma$ upper-limit
\begin{align} \label{YY-R2}
\frac{\big| \raisebox{0pt}{\small$\sum$}_x^{} \texttt Y_{2,1x}^{\textsc{rl}}
\texttt Y_{2,2x}^{\textsc{rl}*} \big|}{m_{R_2}^{}}
& <\, \frac{1.6\times10^{-2}}{\rm TeV} \,. ~~~~~
\end{align}
Barring cancellations among the terms in the summation over $x$, for \,$m_{R_2}=2$ TeV\, this
corresponds to \,$|\textsc k_{\ell_x\ell_x}| < 2.4\times10^{-4}$,\, which is more stringent than
\,$|\textsc k_{ee,\mu\mu,\tau\tau}|\raisebox{1pt}{\footnotesize\,$\lesssim$\,}
(4,2,7)\times10^{-3}$\,
at 95\% CL estimated from data on the high-invariant-mass tails of the dilepton reactions
\,$pp\to\ell^+\ell^-$\, at the LHC and than weaker bounds from
\,$D\to\pi e^+e^-,\pi\mu^+\mu^-$\, measurements~\cite{Fuentes-Martin:2020lea}.

Interestingly, we notice that the condition in Eq.\,(\ref{YY-R2}) no longer matters if the nonzero
elements of the first and second rows of the $\texttt Y_2^{\textsc{rl}}$ matrix do not share same
columns~\cite{Su:2019tjn}, implying that \,$x\neq y$\, in $\textsc k_{\ell_x\ell_y}$.
However, in that case there are restrictions inferred from quests for flavor-violating
\,$pp\to\ell^+\ell^{\prime-}$\, at the LHC, namely
$\big(|\textsc k_{e\mu,e\tau,\mu\tau}|^2+|\textsc k_{\mu e,\tau e,\tau\mu}|^2\big){}^{1/2}
< (2.0,5.8,6.4)\times10^{-3}$\,
at the 2$\sigma$ level~\cite{Angelescu:2020uug}, the first of which is stronger than
\,$|\textsc k_{e\mu,\mu e}|\raisebox{1pt}{\footnotesize\,$\lesssim$\,}(0.01,0.009)$\,
from hunts for rare semileptonic $D_{(s)}$ decays manifesting lepton-flavor violation (LFV),
as discussed in the Appendix.

Accordingly, we can suppose that the only nonvanishing couplings are $\textsc k_{e\mu,\tau\mu}$
and demand that they comply with \,$|\textsc k_{e\mu}|<2.0\times10^{-3}$\, and
\,$|\textsc k_{\tau\mu}|<6.4\times10^{-3}$.\,
This can be realized with a Yukawa matrix having the texture
\begin{align} \label{Y2}
\texttt Y_2^{\textsc{rl}} & \,= \left(\begin{array}{ccc} 0 & \texttt y_{u\mu}^{} & 0 \medskip  \\
\texttt y_{ce}^{} & 0 & \texttt y_{c\tau}^{} \medskip \\ 0 & 0 & 0 \end{array}\right) , ~~~~ ~~~
\end{align}
which can satisfy the limitations from other LFV searches~\cite{Angelescu:2020uug}
and escapes the $D$-mixing restraint.
To see the consequences for \,$c\to u\slashed E$\, explicitly, we incorporate the above
$|\textsc k_{e\mu,\tau\mu}|$ into the expressions displayed in Eqs.\,\,(\ref{BD2Pff'}) and
(\ref{Lc2pff'}) in the Appendix for
the branching fractions of various FCNC decays of charmed mesons and baryons with missing energy.
Thus, we obtain
\begin{align} \label{BD2Pff'-R2}
{\cal B}\big(D^+\to\pi^+\slashed E\big)_{R_2}  & \,<\, 1.6\times10^{-6} \,, &
{\cal B}\big(D^+\to\rho^+\slashed E\big)_{R_2} & \,<\, 8.3\times10^{-7} \,, ~~~ ~~~~
\nonumber \\
{\cal B}\big(D^0\to\pi^0\slashed E\big)_{R_2}  & \,<\, 3.2\times10^{-7} \,, &
{\cal B}\big(D^0\to\eta\slashed E\big)_{R_2}   & \,<\, 9.6\times10^{-8} \,,
\nonumber \\
{\cal B}\big(D^0\to\rho^0\slashed E\big)_{R_2} & \,<\, 1.9\times10^{-7} \,, &
{\cal B}\big(D^0\to\omega\slashed E\big)_{R_2} & \,<\, 1.5\times10^{-7} \,,
\nonumber \\
{\cal B}\big(D^0\to\eta'\slashed E\big)_{R_2}  & \,<\, 1.7\times10^{-8} \,,
\nonumber \\
{\cal B}\big(D_s^+\to K^+\slashed E\big)_{R_2}       & \,<\, 7.5\times10^{-7} \,, &
{\cal B}\big(D_s^+\to K^{*+}\slashed E\big)_{R_2}    & \,<\, 4.7\times10^{-7} \,,
\\
\nonumber \\ \label{BB2B'ff'-R2}
{\cal B}\big(\Lambda_c^+\to p\slashed E\big)_{R_2}   & \,<\, 9.3\times10^{-7} \,, &
{\cal B}\big(\Xi_c^+\to\Sigma^+\slashed E\big)_{R_2} & \,<\, 1.1\times10^{-6} \,, ~~~~~
\nonumber \\
{\cal B}\big(\Xi_c^0\to\Sigma^0\slashed E\big)_{R_2} & \,<\, 3.6\times10^{-7} \,.
\end{align}
where each entry is a sum of branching fractions of the modes with $\nu_e\bar\nu_\mu$ and
$\nu_\tau\bar\nu_\mu$ in the final states, the former making up merely about 10\% of the total.
Also, we find that including $\textsc k_{e\tau,\tau e}$ would barely increase
the preceding results because the associated $\texttt Y_2^{\textsc{rl}}$ elements would have
to fulfill other significant requirements.
Moreover, the aforesaid $D$-mixing requisite
\,$|\textsc k_{\ell_x\ell_x}|\raisebox{1pt}{\footnotesize\,$\lesssim$\,}2.4\times10^{-4}$\,
in the lepton-flavor conserving case would translate into numbers that are smaller by roughly
three orders of magnitude.   
Comparing Eqs.\,(\ref{BD2Pff'-R2})-(\ref{BB2B'ff'-R2}) to the sensitivity reach of ongoing and
future experiments addressed in Sec.\,\ref{intro}, we can conclude that these $R_2$-scenario
predictions probably will not be testable anytime soon.

\section{\boldmath$\tilde R_2$ model\label{tR2}}

From the $\tilde R_2$ part of Eq.\,(\ref{Llq}), in the mass basis of the down-type quarks we have
\begin{align} \label{LtildeR2}
{\cal L}_{\tilde R_2}^{} & \,=\, \tilde{\textsc y}_{2,jy}^{\textsc{lr}} \Big(
{\mathscr V}_{kj}^{}\, \overline{{\cal U}_k^{}} \tilde R_2^{2/3} + \overline{{\cal D}_j^{}}
\tilde R_2^{-1/3} \Big) P_R^{} \textsl{\texttt N}_y^{} \,+\, \rm H.c.
\end{align}
where \,${\mathscr V}\equiv{\mathscr V}_{\textsc{ckm}}$\, designates the Cabibbo-Kobayashi-Maskawa
mixing matrix and \,${\cal D}_{1,2,3}=d,s,b$\, refer to the mass eigenstates.
At low energies, from ${\cal L}_{\tilde R_2}$ proceed the \,$|\Delta C|=1$\, effective
four-fermion interactions specified by
\begin{align} \label{ucNN'-tildeR2}
{\cal L}_{qq'\textsl{\texttt{NN}}'}^{} & \,=\, -\sqrt2\, G_{\rm F}^{}
\Big( \textsc k_{\textsl{\texttt N}_x\textsl{\texttt N}_y}^{\widetilde R_2}\,
\overline u\gamma_\beta^{}P_L^{}c + \widetilde{\textsc k}_{jkxy}^{}\,
\overline{{\cal D}_j^{}}\gamma_\beta^{}P_L^{} {\cal D}_k^{}
\Big) \overline{\textsl{\texttt N}_x^{}} \gamma^\beta P_R^{} \textsl{\texttt N}_y^{}
\,+\, {\rm H.c.} \,, ~~~~~
\end{align}
where \,$j,k,x,y=1,2,3$\, are implicitly summed over,
\begin{align} \label{Cff'}
\textsc k_{\textsl{\texttt N}_x\textsl{\texttt N}_y}^{\tilde R_2} & =\,
\frac{\hat v^2 ({\mathscr V}\tilde{\textsc y}{}_2^{\textsc{lr}})_{1y}^{}
\big({\mathscr V}\tilde{\textsc y}{}_2^{\textsc{lr}})_{2x}^*}{2 m_{\tilde R_2}^2} \,, &
\widetilde{\textsc k}_{jkxy}^{} & =\, \frac{\hat v^2\, \tilde{\textsc y}{}_{2,jy}^{\textsc{lr}}
\tilde{\textsc y}{}_{2,kx}^{\textsc{lr}*} }
{2 m_{\tilde R_2}^2} \,. ~~~ ~~~~
\end{align}
As in the last section, this brings about the FCNC decays of charmed hadrons with missing energy,
but now it is the right-handed neutrinos that act as the invisibles.
Furthermore, ${\cal L}_{qq'\textsl{\texttt{NN}}'}$ can induce analogous transitions among
down-type quarks.

From Eq.\,(\ref{LtildeR2}), one can calculate the box diagrams, with $\tilde R_2$ and
$\textsl{\texttt N}_y$ running around the loops, which affects $D^0$-$\bar D^0$ mixing,
like in the $R_2$ scenario.
This leads to the effective Hamiltonian
\begin{align} \label{Hucuc-tildeR2}
{\cal H}_{|\Delta C|=2}^{\tilde R_2} & \,=\, \frac{ \Big[ \mbox{$\sum$}_x^{}
\big({\mathscr V}\tilde{\textsc y}{}_2^{\textsc{lr}}\big)_{1x}
\big({\mathscr V}\tilde{\textsc y}{}_2^{\textsc{lr}}\big)_{2x}^* \Big]\raisebox{5pt}{$^2$}}
{128\pi^2\, m_{\tilde R_2}^2}\, \overline u\gamma^\beta P_L^{}c\,\overline u\gamma_\beta^{}P_L^{}c
\,+\, {\rm H.c.} ~~~ ~~~~
\end{align}
However, differently from before, there are additionally contributions to its kaon and $b$-meson
($B_d$ and $B_s$) counterparts, described by
\begin{align} \label{Hsdsd-tildeR2}
{\cal H}_{|\Delta S|=2}^{\tilde R_2} & \,=\, \frac{\big( \mbox{$\sum$}_x^{}
\tilde{\textsc y}{}_{2,2x}^{\textsc{lr}} \tilde{\textsc y}{}_{2,1x}^{\textsc{lr}*}
\big)\raisebox{1pt}{$\!^2$}}{128\pi^2\, m_{\tilde R_2}^2}\, \overline s\gamma^\beta P_L^{}d\,
\overline s\gamma_\beta^{}P_L^{}d \,+\, {\rm H.c.} ~~~ ~~~~
\end{align}
and similar formulas in the $B_{d,s}$-mixing cases.
Since we cannot avoid all the mixing constraints at the same time, we can opt instead to do so in
the down-type sector alone.
It is evident from Eq.\,(\ref{Hsdsd-tildeR2}) that this is attainable if the nonzero
elements of $\tilde{\textsc y}{}_2^{\textsc{lr}}$ lie in separate columns. 
Accordingly, for simplicity we can pick
\begin{align} \label{tY2}
\tilde{\textsc y}{}_2^{\textsc{lr}} & \,= \left(\begin{array}{ccc} 0 & 0 & 0 \medskip \\
0 & ~0~ & \tilde{\texttt y}_{s3}^{} \medskip \\ 0 & 0 & 0 \end{array}\right) , ~~~ ~~~~
\end{align}
and so we have
\begin{align} \label{VtY2}
{\mathscr V}\tilde{\textsc y}{}_2^{\textsc{lr}} & \,= \left(\begin{array}{ccc} 0 & 0 & \lambda
\medskip \\ 0 & ~0~ & 1-\tfrac{1}{2}\lambda^2 \medskip \\
0 & 0 & -\lambda^2 A \end{array}\right) \tilde{\texttt y}_{s3} \,, ~~~ ~~~~
\end{align}
up to ${\cal O}(\lambda^2)$, where $\lambda$ and $A$ are Wolfenstein parameters.
To this order in \,$\lambda\sim0.23$\, the nonvanishing coefficients in Eq.\,(\ref{Cff'}) are
\begin{align} \label{KK}
\textsc k_{\textsl{\texttt N}_3\textsl{\texttt N}_3}^{\tilde R_2} &\, =\, \frac{\hat v^2
({\mathscr V}\tilde{\textsc y}{}_2^{\textsc{lr}})_{13}^{}
\big({\mathscr V}\tilde{\textsc y}{}_2^{\textsc{lr}})_{23}^*}{2 m_{\tilde R_2}^2}
\,=\, \frac{\lambda\, \hat v^2 |\tilde{\texttt y}_{s3}|^2}{2 m_{\tilde R_2}^2} \,, ~~~ ~~~~
\nonumber \\
\widetilde{\textsc k}_{2233}^{} & \,=\, \frac{\hat v^2\,
|\tilde{\textsc y}{}_{2,23}^{\textsc{lr}}|^2}{2 m_{\tilde R_2}^2}
\,=\, \frac{\hat v^2\, |\tilde{\texttt y}_{s3}|^2}{2 m_{\tilde R_2}^2} \,.
\end{align}

Using
\,$\langle\bar D^0|\overline u\gamma^\kappa P_Lc \overline u\gamma_\kappa^{}P_Lc|D^0\rangle =
0.0805(57)$\,GeV$^4$\,
from lattice QCD work~\cite{Bazavov:2017weg} and demanding again that the $\tilde R_2$
contribution saturate $\Delta m_D^{\rm exp}$, we get
\begin{align} \label{YY-tildeR2}
\frac{ \Big| \mbox{$\sum$}_x^{} \big({\mathscr V}\tilde{\textsc y}{}_2^{\textsc{lr}}\big)_{1x}
\big({\mathscr V}\tilde{\textsc y}{}_2^{\textsc{lr}}\big)_{2x}^* \Big| }{m_{\tilde R_2}^{}}
& <\, \sqrt{\frac{128\pi^2 \Delta m_D^{\rm exp}\, m_{D^0}^{}}{\langle\bar D^0|
\overline u\gamma^\kappa P_Lc\, \overline u\gamma_\kappa^{}P_Lc|D^0\rangle \tilde r} }
\,=\, \frac{2.3\times10^{-2}}{\rm TeV} ~~~~ ~~~
\end{align}
at the 2$\sigma$ level.
For \,$m_{\tilde R_2}=2$\,TeV,\, this translates into
\begin{align} \label{KNN'-tR2}
\big|\textsc k_{\textsl{\texttt N}_3\textsl{\texttt N}_3}^{\tilde R_2}\big| & \,<\,
3.4\times10^{-4} \,. ~~~~ ~~~
\end{align}
With this coupling value, the charmed-hadron decay channels with missing energy listed in
Eqs.\,\,(\ref{BD2Pff'-R2}) and (\ref{BB2B'ff'-R2}) would turn out to have branching fractions
about two orders of magnitude smaller than the corresponding numbers displayed therein.
We further find that having more nonzero elements in, say, the first two rows of
$\tilde{\textsc y}{}_2^{\textsc{lr}}$ would produce little change to this conclusion
because they would be subject mainly to the meson-mixing requisites and/or stringent bounds
inferred from \,$K\to\pi\slashed E$\, measurements.

\section{\boldmath$\bar S_1$ model\label{bS1}}

From the $\bar S_1$ portion of Eq.\,(\ref{Llq}),
\begin{align} \label{LbarS1}
{\cal L}_{\bar S_1}^{} & \,=\, \bar{\textsc y}_{1,jy}^{\textsc{rr}}\,
\overline{{\cal U}_j^{\textsc c}} P_R^{} \textsl{\texttt N}_y^{} \bar S_1^{-2/3}
\,+\, \rm H.c. \,, ~~~~~
\end{align}
we derive
\begin{align} \label{ucNN'-barS1}
{\cal L}_{uc\textsl{\texttt{NN}}'}^{} & \,=\, -\sqrt2\, G_{\rm F}^{}\,
\textsc k_{\textsl{\texttt N}_x\textsl{\texttt N}_y}^{\bar S_1}
\overline u\gamma^\beta P_R^{}c\, \overline{\textsl{\texttt N}_x^{}} \gamma_\beta^{} P_R^{}
\textsl{\texttt N}_y^{} \,+\, {\rm H.c.} \,, ~~ ~~~
\end{align}
where
\begin{align} \label{KNN'-bS1}
\textsc k_{\textsl{\texttt N}_x\textsl{\texttt N}_y}^{\bar S_1} & \,=\,
\frac{-\hat v^2\, \bar{\textsc y}{}_{1,1x\,}^{\textsc{rr}*}\bar{\textsc y}{}_{1,2y}^{\textsc{rr}}}
{2 m_{\bar S_1}^2} \,. ~~ ~~~
\end{align}
This again gives rise to \,$c\to u\textsl{\texttt N}_x\bar{\textsl{\texttt N}}{}_y$, with
$\textsl{\texttt N}_x\bar{\textsl{\texttt N}}{}_y$ emitted invisibly, and affects $\Delta m_D$,
the latter via
\begin{align} \label{Hucuc-barS1}
{\cal H}_{|\Delta C|=2}^{\bar S_1} & \,=\, \frac{ \big( \mbox{$\sum$}_x^{}
\bar{\textsc y}{}_{1,1x}^{\textsc{rr}*} \bar{\textsc y}{}_{1,2x}^{\textsc{rr}}
\big)\raisebox{1pt}{$\!^2$} }{128\pi^2\, m_{\bar S_1}^2}\, \overline u\gamma^\beta P_R^{}c\,
\overline u\gamma_\beta^{}P_R^{}c  \,+\, {\rm H.c.} ~~~~~
\end{align}
Hence the mixing requirement is escapable if the contributing elements of the first and second
rows of $\bar{\textsc y}{}_1^{\textsc{rr}}$ belong to different columns, as in this simple example:
\begin{align} \label{bY1}
\bar{\textsc y}{}_1^{\textsc{rr}} & \,= \left(\begin{array}{ccc} 0 & \bar{\texttt y}_{u2} & 0
\medskip \\ \bar{\texttt y}_{c1} & 0 & \bar{\texttt y}_{c3} \medskip \\
0 & 0 & 0 \end{array}\right) . ~~~~ ~~~
\end{align}

With \,$x\neq y$\, in $\textsc k_{\textsl{\texttt N}_x\textsl{\texttt N}_y}^{\bar S_1}$, 
the remaining consequential limitation on the $\bar{\textsc y}{}_1^{\textsc{rr}}$ elements is
that from the perturbativity condition:
\,$|\bar{\textsc y}{}_{1,ix}^{\textsc{rr}}|<\sqrt{4\pi}$.\,
As for the allowed range of the $\bar S_1$ mass, the latest quest by the CMS Collaboration
\cite{Sirunyan:2018kzh} for scalar LQs decaying fully into a quark and neutrino has ruled out
masses up to 1.1 TeV at 95\% CL.
Since this is applicable to the possibility that the neutrino is a right-handed one,
we can set \,$m_{\bar S_1} > 1.2$\,TeV.\,
These parameters also enter loop diagrams involving $\bar S_1$ and the $u$ and $c$ quarks and
modifying the invisible partial width of the $Z$ boson, but we have checked that their impact
is insignificant.
Incorporating these numbers into Eq.\,(\ref{KNN'-bS1}) yields, for \,$x\neq y$,\,
\begin{align} \label{Kmax}
\big|\textsc k_{\textsl{\texttt N}_x\textsl{\texttt N}_y}^{\bar S_1}\big| & \,<\,
0.26 \,. ~~~~ ~~~
\end{align}

To illustrate the implications for the aforementioned charmed-hadron decays, we adopt
the Yukawa matrix in Eq.\,(\ref{bY1}), in which case only
$\textsc k_{\textsl{\texttt N}_2\textsl{\texttt N}_1}^{\bar S_1}$ and
$\textsc k_{\textsl{\texttt N}_2\textsl{\texttt N}_3}^{\bar S_1}$ are present.
Assuming that they each have the maximal value in Eq.\,(\ref{Kmax}) and putting them together 
with Eqs.\,\,(\ref{BD2Pff'}) and (\ref{Lc2pff'}), we then arrive at
\begin{align} \label{BD2Pff'-barS1}
{\cal B}\big(D^+\to\pi^+\slashed E\big)_{\bar S_1}  & \,<\, 4.9\times10^{-3} \,, &
{\cal B}\big(D^+\to\rho^+\slashed E\big)_{\bar S_1} & \,<\, 2.5\times10^{-3} \,, ~~~~~
\nonumber \\
{\cal B}\big(D^0\to\pi^0\slashed E\big)_{\bar S_1}  & \,<\, 9.7\times10^{-4} \,, &
{\cal B}\big(D^0\to\eta\slashed E\big)_{\bar S_1}   & \,<\, 2.9\times10^{-4} \,,
\nonumber \\
{\cal B}\big(D^0\to\rho^0\slashed E\big)_{\bar S_1} & \,<\, 5.7\times10^{-4} \,, &
{\cal B}\big(D^0\to\omega\slashed E\big)_{\bar S_1} & \,<\, 4.4\times10^{-4} \,,
\nonumber \\
{\cal B}\big(D^0\to\eta'\slashed E\big)_{\bar S_1}  & \,<\, 5.2\times10^{-5} \,,
\nonumber \\
{\cal B}\big(D_s^+\to K^+\slashed E\big)_{\bar S_1}       & \,<\, 2.2\times10^{-3} \,, &
{\cal B}\big(D_s^+\to K^{*+}\slashed E\big)_{\bar S_1}    & \,<\, 1.4\times10^{-3} \,,
\\
\nonumber \\ \label{BB2B'ff'-barS1}
{\cal B}\big(\Lambda_c^+\to p\slashed E\big)_{\bar S_1}   & \,<\, 2.8\times10^{-3} \,, &
{\cal B}\big(\Xi_c^+\to\Sigma^+\slashed E\big)_{\bar S_1} & \,<\, 3.2\times10^{-3} \,, ~~
\nonumber \\
{\cal B}\big(\Xi_c^0\to\Sigma^0\slashed E\big)_{\bar S_1} & \,<\, 1.1\times10^{-3} \,,
\end{align}
where each entry is a combination of branching fractions of the modes with
$\textsl{\texttt N}_2\bar{\textsl{\texttt N}}{}_1$ and
$\textsl{\texttt N}_2\bar{\textsl{\texttt N}}{}_3$ carrying away the missing energy 
in the final states.
These numbers are considerably higher than their counterparts in the models containing $R_2$
and $\tilde R_2$.
This is attributable to the fact that $\bar S_1$ does not have any direct couplings to
the SM lepton and quark doublets.
Additionally, one can see that some of the results in 
Eqs.\,(\ref{BD2Pff'-barS1})-(\ref{BB2B'ff'-barS1}) are within the sensitivity reach of BESIII 
and Belle II described in Sec.\,\ref{intro}, suggesting that they might soon discover one or more 
of these $\bar S_1$-mediated processes or, if not, set useful bounds on them.

It is worth noting that our selection above for the Yukawa couplings of $\bar S_1$ can be 
explained in terms of flavor symmetry imposed on the interactions of the sterile neutrinos.
Specifically, supposing that $\textsl{\texttt N}_2^\dagger$ and $\textsl{\texttt N}_{1,3}^\dagger$
carry, respectively, what may be called ``upness'' and ``charmness'' quantum numbers associated 
with the right-handed mass-eigenstates of the $u$ and $c$ quarks, we can see that 
${\cal L}_{\bar S_1}$ in Eq.\,(\ref{LbarS1}) with $\bar{\textsc y}{}_1^{\textsc{rr}}$ picked to be 
of the form in Eq.\,(\ref{bY1}) conserves these numbers, as do 
\,$c\to u\textsl{\texttt N}_2^{}\bar{\textsl{\texttt N}}{}_1^{}$\, and 
\,$c\to u\textsl{\texttt N}_2^{}\bar{\textsl{\texttt N}}{}_3^{}$\, following from it.\footnote{Here 
we have taken the light $\textsl{\texttt N}_1$ and $\textsl{\texttt N}_3$ to have unequal masses.
If these are the same instead, the $\textsl{\texttt N}_{1,3}$ fields can be rotated such that only 
one of their linear combinations participates in \,$c\to u\slashed E$\, 
along with $\textsl{\texttt N}_2$.}
At the same time, this choice prevents $\textsl{\texttt N}_{1,2,3}$ from affecting 
$D^0$-$\bar D^0$ mixing via the Hamiltonian in Eq.\,(\ref{Hucuc-barS1}),
which violates the symmetry.

\section{Conclusions\label{concl}}

We have explored the FCNC decays of charmed hadrons into a lighter hadron and missing energy
carried away by a pair of either SM or right-handed sterile neutrinos in LQ scenarios,
concentrating on the influence of the $R_2$, $\tilde R_2$, and $\bar S_1$ scalar LQs.
We take into account various relevant constraints and learn that the meson-mixing ones and
those inferred from LHC searches are especially important.
Nevertheless, we point out that the meson-mixing restrictions may be evaded in certain situations.
Additionally, we demonstrate that the contributions of these LQs to the branching fractions of
\,$D^+\to\mathscr M^+\slashed E$, \,$\mathscr M=\pi,\rho$,\, of
\,$D^0\to\tilde{\mathscr M}\slashed E$, \,$\tilde{\mathscr M}=\pi^0,\eta,\rho^0,\omega,\eta'$,\,
and of \,$D_s^+\to K^{(*)+}\slashed E$\, can be evaluated without knowing the details of the mesonic
form factors associated with the quark currents if the invisibles have vanishing masses,
by employing the data on the corresponding semileptonic modes and assuming isospin symmetry.
As a consequence, the calculated $D_{(s)}$ rates are free from the uncertainties attendant in
form-factor estimation.

Our numerical work indicates that several of these charmed-meson decays and their baryon
counterparts \,$\Lambda_c^+\to p\slashed E$\, and \,$\Xi_c^{+,0}\to\Sigma^{+,0}\slashed E$\, are
currently permitted to have branching fractions reaching the $10^{-7}$-$10^{-6}$ levels if $R_2$
is responsible for the underlying operators.
On the other hand, the effects of $\tilde R_2$ are comparatively much less.
By contrast, the contributions of $\bar S_1$, which has fermionic couplings exclusively to
right-handed up-type quarks and the sterile neutrinos, could produce substantially bigger
branching fractions, up to a few times $10^{-3}$.
This is understandable because $R_2$ and $\tilde R_2$ are directly linked to the SM
left-handed lepton and quark doublets, respectively, implying relatively stronger restraints on
the Yukawa-matrix elements of these LQs.
We therefore conclude that the charmed hadron decays we have studied are potentially promising as
probes of LQ interactions involving sterile neutrinos.
Lastly, a number of our predictions, notably in the $\bar S_1$ scenario, may be large enough to be
testable in the ongoing BESIII and Belle II experiments.

\acknowledgments

This research was supported in part by the MOST (Grant No. MOST 106-2112-M-002-003-MY3).

\appendix

\section{Branching ratio formulas\label{Brelations}}

The effective Lagrangian for the \,$c\to u$\, transitions of interest has the form
\begin{align} \label{Lucff'}
{\cal L}_{uc\texttt{ff}'} & \,=\, -\overline u\gamma^\kappa c\;
\overline{\texttt f} \gamma_\kappa^{} \big( \texttt C_{\texttt{ff}'}^{\textsc v} + \gamma_5^{}
\texttt C_{\texttt{ff}'}^{\textsc a} \big) \texttt f' - \overline u\gamma^\kappa\gamma_5^{}c\;
\overline{\texttt f} \gamma_\kappa^{} \big( \tilde{\textsf c}{}_{\texttt{ff}'}^{\textsc v}
+ \gamma_5^{} \tilde{\textsf c}{}_{\texttt{ff}'}^{\textsc a} \big) \texttt f' \,+\, {\rm H.c.} \,,
\end{align}
where \texttt f and \texttt f$'$ are either SM leptons or SM-gauge-singlet fermions.
This gives rise to \,$D\to{\mathbb P}\texttt f\bar{\texttt f}{}'$\, and
\,$D\to{\mathbb V}\texttt f\bar{\texttt f}{}'$,\, where $D$ stands for a  charmed pseudoscalar
meson and $\mathbb P$ and $\mathbb V$ designate charmless pseudoscalar and vector mesons,
respectively.
The amplitudes ${\cal M}_{D\to{\mathbb P}\texttt f\bar{\texttt f}{}'}$ and
${\cal M}_{D\to{\mathbb V}\texttt f\bar{\texttt f}{}'}$ for these decays depend on
the mesonic matrix elements~\cite{Dorsner:2016wpm,Isgur:1990kf}
\begin{align} \label{D->P}
\langle{\mathbb P}|\overline u\gamma^\alpha c|D\rangle & \,=\,
F_+^{}\, \tilde p^\alpha + F_-^{}\, \tilde q^\alpha \,, ~~~ ~~~~
\langle{\mathbb V}|\overline u\gamma^\alpha c|D\rangle \,=\, \frac{V}{\tilde{\textsl{\texttt m}}_+}\,
\epsilon^{\alpha\rho\sigma\tau}\, \varepsilon_\rho^* \tilde q_\sigma^{}\tilde p_\tau^{} \,,
\nonumber \\
\langle{\mathbb V}|\overline u\gamma^\alpha\gamma_5^{}c|D\rangle & \,=\, i \bigg[ \frac{2 A_0^{}
m_{\mathbb V}^{}}{\tilde q^2} \tilde q^\alpha - \frac{A_2^{}}{\tilde{\textsl{\texttt m}}_+}
\bigg( \tilde p^\alpha - \frac{\tilde{\textsl{\texttt m}}_+^{}\tilde{\textsl{\texttt m}}_-^{}}{\tilde q^2}
\tilde q^\alpha \bigg) \bigg] \varepsilon_\kappa^* \tilde q^\kappa
+ i A_1^{} \tilde{\textsl{\texttt m}}_+^{} \bigg( \varepsilon^{*\alpha}
- \frac{\varepsilon_\kappa^*\tilde q^\kappa}{\tilde q^2}\tilde q^\alpha \bigg) \,, ~~~
\end{align}
where $m_X^{}$ and $p_X^{}$ are the mass and momentum of $X$, respectively,
\begin{align}
\tilde{\textsl{\texttt m}}_\pm^{} & \,=\, m_D^{}\pm m_{\mathbb V}^{} \,, &
\tilde p & \,=\, p_D^{}+p_{\mathbb M}^{} \,, & \tilde q & \,=\, p_D^{}-p_{\mathbb M}^{} \,, &
\mathbb M & \,=\, \mathbb P, \mathbb V \,, ~~~
\end{align}
$\varepsilon$ denotes the polarization vector of $\mathbb V$, and $F_\pm^{}$, $V$, and
$A_{0,1,2}$ symbolize form factors which are functions of $\tilde q^2$.
In this paper, we focus on the possibility that the \texttt f and \texttt f$'$ masses,
$m_{\texttt f}$ and $m_{\texttt f'}$, are sufficiently small to be negligible.\footnote{In that
case \,$D^0\to\texttt f\bar{\texttt f}{}'$\, arising from ${\cal L}_{uc\texttt{ff}'}$ is chirally
suppressed and hence has a tiny rate.}
It follows that
\begin{align} \label{MB2Pff'}
{\cal M}_{D\to{\mathbb P}\texttt f\bar{\texttt f}{}'}^{} & \,=\, 2 F_+^{}\, \bar u_{\texttt f}^{}\,
\slashed p_{\mathbb P} \big( \texttt C_{\texttt{ff}'}^{\textsc v} + \gamma_5^{}
\texttt C_{\texttt{ff}'}^{\textsc a} \big) v_{\bar{\texttt f}{}'}^{} \,, &
\end{align}
\begin{align} \label{MB2Vff'}
{\cal M}_{D\to{\mathbb V}\texttt f\bar{\texttt f}{}'}^{} & \,=\, i \bigg( A_1^{}
\tilde{\textsl{\texttt m}}_+^{} \varepsilon_\alpha^* - \frac{A_2}{\tilde{\textsl{\texttt m}}_+}
\varepsilon_\kappa^* \tilde q^\kappa \tilde p_\alpha^{} \bigg) \bar u_{\texttt f}^{}
\gamma^\alpha \big( \tilde{\textsf c}{}_{\texttt{ff}'}^{\textsc v} + \gamma_5^{}
\tilde{\textsf c}{}_{\texttt{ff}'}^{\textsc a} \big) v_{\bar{\texttt f}{}'}^{}
\nonumber \\ & ~~~~ +\,
\frac{V}{\tilde{\textsl{\texttt m}}_+}\, \epsilon^{\alpha\rho\sigma\tau} \varepsilon_\rho^*
\tilde q_\sigma^{} \tilde p_\tau^{}\, \bar u_{\texttt f}^{} \gamma_\alpha^{} \big(
\texttt C_{\texttt{ff}'}^{\textsc v} + \gamma_5^{} \texttt C_{\texttt{ff}'}^{\textsc a} \big)
v_{\bar{\texttt f}{}'}^{} \,, &
\end{align}
where $u_{\texttt f}^{}$ and $v_{\bar{\texttt f}{}'}$ represent the fermions' Dirac spinors and
the contributions of the terms with \,$\tilde q^\alpha=p_{\texttt f}^\alpha+p_{\texttt f'}^\alpha$\,
in Eq.\,(\ref{D->P}) have dropped out upon contraction with the $\texttt f\bar{\texttt f}{}'$
current due to \,$m_{\texttt f,\texttt f'}\simeq0$.
For \,$\texttt f\neq\texttt f'$,\, these amplitudes translate into the differential rates
\begin{align} \label{G'B2Pff'}
\frac{d\Gamma_{D\to{\mathbb P}\texttt f\bar{\texttt f}{}'}^{}}{d\hat s} & \,=\,
\frac{\tilde\lambda_{D\mathbb P}^{3/2}\, F_+^2}{192\pi^3 m_D^3}
\big(|\texttt C_{\texttt{ff}'}^{\textsc v}|^2+|\texttt C_{\texttt{ff}'}^{\textsc a}|^2\big) \,, &
\end{align}
\begin{align} \label{G'B2Vff'}
\frac{d\Gamma_{D\to{\mathbb V}\texttt f\bar{\texttt f}{}'}}{d\hat s} \,=\,
\frac{\tilde\lambda_{D\mathbb V}^{3/2}}{768 \pi^3 m_D^3} & \Bigg\{ \Bigg[ A_1^2\, \tilde{\textsl{\texttt m}}{}_+^2
\Bigg( 1 + \frac{12 m_{\mathbb V}^2\hat s}{\tilde\lambda_{D\mathbb V}^{}} \Bigg)
+ \tilde\varsigma A_1^{}A_2^{} + \frac{\tilde\lambda_{D\mathbb V}^{} A_2^2}{\tilde{\textsl{\texttt m}}{}_+^2}
\Bigg] \frac{ |\tilde{\textsf c}{}_{\texttt{ff}'}^{\textsc v}|^2
+ |\tilde{\textsf c}{}_{\texttt{ff}'}^{\textsc a}|^2 }{m_{\mathbb V}^2}  ~~~~~
\nonumber \\ & ~+\,
\frac{8 V^2 \hat s}{\tilde{\textsl{\texttt m}}{}_+^2} \big( |\texttt C_{\texttt{ff}'}^{\textsc v}|^2
+ |\texttt C_{\texttt{ff}'}^{\textsc a}|^2 \big) \Bigg\} \,,
\end{align}
where
\begin{align}
\hat s & \,=\, (p_{\texttt f}^{}+p_{\texttt f'}^{})^2 \,, &
\tilde\lambda_{XY}^{} & \,=\, \big(m_X^2-m_Y^2-\hat s\big){}^2-4m_Y^2\hat s \,, &
\tilde\varsigma & \,=\, 2\hat s-2 \tilde{\textsl{\texttt m}}_+^{} \tilde{\textsl{\texttt m}}_-^{} \,. ~~~
\end{align}
For the LQ-mediated operators in Eqs.\,\,(\ref{ucff'-R2}), (\ref{ucNN'-tildeR2}), and
(\ref{ucNN'-barS1}) containing constants of the form $\textsc k_{\texttt{ff}'}$, we can then
apply Eqs.\,\,(\ref{G'B2Pff'}) and (\ref{G'B2Vff'}) by setting
\,$|\texttt C_{\texttt{ff}'}^{\textsc v,\textsc a}| =
|\tilde{\textsf c}{}_{\texttt{ff}'}^{\textsc v,\textsc a}| =
G_{\rm F} |\textsc k_{\texttt{ff}'}|/\sqrt8$.

Now, the semileptonic transitions \,$D^0\to\mathscr M^-\nu e^+$\, with \,$\mathscr M=\pi, \rho$\,
receive SM contributions described by
\,${\cal L}_{dc\nu e}^{\textsc{sm}} = -\sqrt8\,G_{\rm F}^{}V_{cd}^*\,
\overline d\gamma^\alpha P_L^{}c\, \overline{\nu_e^{}}\gamma_\alpha^{}P_L^{}e + {\rm H.c.}$\,
Comparing this to ${\cal L}_{uc\texttt{ff}'}$ in Eq.\,(\ref{Lucff'}) and ignoring the leptons'
masses, we can see that the expressions for the differential rates of $D^0\to\mathscr M^-\nu e^+$\,
with \,$\mathscr M=\pi, \rho$\, in the SM are equal to those in Eqs.\,\,(\ref{G'B2Pff'}) and
(\ref{G'B2Vff'}), respectively, but with coefficients given by
\,$|\texttt C_{\nu e}^{\textsc v,\textsc a}| = |\tilde{\textsf c}{}_{\nu e}^{\textsc v,\textsc a}|
= G_{\rm F} |V_{cd}|/\sqrt2$.\,
For the rate of LQ-induced \,$D^+\to\mathscr M^+\texttt f\bar{\texttt f}{}'$, neglecting small
isospin-breaking effects we then arrive at
\,$4 \Gamma_{D^+\to\mathscr M^+\texttt f\bar{\texttt f}{}'}^{\textsc{lq}} |V_{cd}|^2 =
\Gamma_{D^0\to\mathscr M^-\nu e^+}^{\textsc{sm}} |\textsc k_{\texttt{ff}'}|^2$\,
without having to know how $F_+^{}$, $V$, and $A_{1,2}$ depend on $\hat s$.
As the LQ interactions in Eq.\,(\ref{Llq}) do not directly affect \,$D^0\to\mathscr M^-\nu e^+$,\,
we can replace $\Gamma_{D^0\to\mathscr M^-\nu e^+}^{\textsc{sm}}$ with their experimental values.
This implies the branching-fraction relation
\begin{align} \label{D2Mff'}
{\cal B}(D^+\to\mathscr M^+\texttt f\bar{\texttt f}{}')_{\textsc{lq}}^{}  & \,=\,
\frac{\tau_{D^+}^{}}{\tau_{D^0}^{}} \frac{{\cal B}(D^0\to\mathscr M^-\nu e^+)_{\rm exp}^{}}
{4 |V_{cd}|^2}\, |\textsc k_{\texttt{ff}'}|^2 \,, ~~~~~
\end{align}
where $\tau_{D^{+(0)}}$ is the measured $D^{+(0)}$ lifetime.
It is straightforward to write down analogous formulas for other modes, particularly
\,$D^0\to\tilde{\mathscr M}\texttt f\bar{\texttt f}{}'$,
\,$\tilde{\mathscr M}=\pi^0,\eta,\rho^0,\omega,\eta'$,\, and
\,$D_s^+\to K^{(*)+}\texttt f\bar{\texttt f}{}'$.\,
Clearly, the outcomes of this procedure do not suffer from the uncertainties inherent in
the estimation of hadronic matrix elements.

To proceed, we need the empirical information on the relevant semileptonic
modes \cite{Zyla:2020zbs}:
\begin{align} \label{xBD2Pln}
{\cal B}(D^0\to\pi^-\nu e^+)_{\rm exp}^{} & \,=\, 2.91(4) \,, &
{\cal B}(D^0\to\rho^-\nu e^+)_{\rm exp}^{} & \,=\, 1.50(12) \,,
\nonumber \\
{\cal B}(D^+\to\pi^0\nu e^+)_{\rm exp}^{} & \,=\, 3.72(17) \,, &
{\cal B}(D^+\to\eta\nu e^+)_{\rm exp}^{} & \,=\, 1.11(7) \,,
\nonumber \\
{\cal B}(D^+\to\rho^0\nu e^+)_{\rm exp}^{} & \,=\, 2.18^{+0.17}_{-0.25} \,, &
{\cal B}(D^+\to\omega\nu e^+)_{\rm exp}^{} & \,=\, 1.69(11) \,,
\nonumber \\
{\cal B}(D^+\to\eta'\nu e^+)_{\rm exp}^{} & \,=\, 0.20(4) \,,
\nonumber \\
{\cal B}\big(D_s^+\to K^0\nu e^+\big)_{\rm exp} & \,=\, 3.4(4) \,, &
{\cal B}\big(D_s^+\to K^{*0}\nu e^+\big)_{\rm exp} & \,=\, 2.15(28) ~~~~
\end{align}
all in units of $10^{-3}$.
Using their central values and the CKM matrix element
\,$|V_{cd}|=0.22636(48)$~\cite{Zyla:2020zbs}, we then find
\begin{align} \label{BD2Pff'}
{\cal B}(D^+\to\pi^+\texttt f\bar{\texttt f}{}')_{\textsc{lq}}^{} & \,=\,
3.60 \times 10^{-2}\, |\textsc k_{\texttt{ff}'}|^2 \,, &
\nonumber \\
{\cal B}\big(D^+\to\rho^+\texttt f\bar{\texttt f}{}'\big)_{\textsc{lq}} & \,=\,
1.86 \times 10^{-2}\, |\textsc k_{\texttt{ff}'}|^2 \,,
\nonumber \\
\nonumber \\
{\cal B}(D^0\to\pi^0\texttt f\bar{\texttt f}{}')_{\textsc{lq}}^{} & \,=\,
7.16 \times 10^{-3}\, |\textsc k_{\texttt{ff}'}|^2 \,,  
\nonumber \\
{\cal B}\big(D^0\to\eta\texttt f\bar{\texttt f}{}'\big)_{\textsc{lq}} & \,=\,
2.14 \times 10^{-3}\, |\textsc k_{\texttt{ff}'}|^2 \,,
\nonumber \\
{\cal B}\big(D^0\to\rho^0\texttt f\bar{\texttt f}{}'\big)_{\textsc{lq}} & \,=\,
4.19 \times 10^{-3}\, |\textsc k_{\texttt{ff}'}|^2 \,,  
\nonumber \\
{\cal B}(D^0\to\omega\texttt f\bar{\texttt f}{}')_{\textsc{lq}}^{} & \,=\,
3.25 \times 10^{-3}\, |\textsc k_{\texttt{ff}'}|^2 \,,
\nonumber \\
{\cal B}\big(D^0\to\eta'\texttt f\bar{\texttt f}{}'\big)_{\textsc{lq}} & \,=\,
3.8 \times 10^{-4}\, |\textsc k_{\texttt{ff}'}|^2 \,,
\nonumber \\
\nonumber \\
{\cal B}\big(D_s^+\to K^+\texttt f\bar{\texttt f}{}'\big)_{\textsc{lq}}    & \,=\,
1.7 \times 10^{-2}\, |\textsc k_{\texttt{ff}'}|^2 \,,  
\nonumber \\
{\cal B}\big(D_s^+\to K^{*+}\texttt f\bar{\texttt f}{}'\big)_{\textsc{lq}} & \,=\,
1.05 \times 10^{-2}\, |\textsc k_{\texttt{ff}'}|^2 \,.
\end{align}
The numbers in Eq.\,(\ref{BD2Pff'}) have relative errors approximately equal to those of
the corresponding data in Eq.\,(\ref{xBD2Pln}).

We can apply the preceding results to extract bounds on $\textsc k_{e\mu,\mu e}$ defined in
Eq.\,(\ref{ucff'-R2}) from hunts for \,$D_{(s)}\to\pi(K)e^\pm\mu^\mp$,\,
the channels with the tau lepton being kinematically closed.
In light of Eq.\,(\ref{BD2Pff'}) and the available limits on the pertinent
modes~\cite{Zyla:2020zbs}, the strongest restraints come from
\,${\cal B}(D^+\to\pi^+e^-\mu^+)_{\rm exp}^{}<3.6\times10^{-6}$\, and
\,${\cal B}(D^+\to\pi^+e^+\mu^-)_{\rm exp}^{}<2.9\times10^{-6}$,\, both at 90\%
CL~\cite{Zyla:2020zbs}, which translate into
\begin{align} \label{Kem}
|\textsc k_{e\mu}| & \,<\, 1.0\times10^{-2} \,, &
|\textsc k_{\mu e}| & \,<\, 9.0\times10^{-3} \,, &
\end{align}
respectively.
These turn out to be less stringent than those implied by \,$D^0\to e^\pm\mu^\mp$ data
\cite{Zyla:2020zbs} and inferred from quests for
\,$pp\to e^\pm\mu^\mp$\, at the LHC \cite{Angelescu:2020uug}.

It is worth mentioning that when using Eq.\,(\ref{BD2Pff'}) for the models in
Secs. \ref{R2}-\ref{bS1}, where \texttt f and \texttt f$'$ are invisible, we ignore the SM
contributions, which are highly suppressed \cite{Golowich:2007ka}.
We also note that \,$D_{(s)}^+\to\mathscr M_{(s)}^+\texttt f\bar{\texttt f}{}'$\, with
\,$\mathscr M_{(s)}=\pi, \rho\, (K,K^*)$\, and invisible \texttt f and \texttt f$'$
have SM backgrounds from the sequential decays
\,$D_{(s)}^+\to\tau^+\nu$\, and \,$\tau^+\to\mathscr M_{(s)}^+\nu$ \cite{Burdman:2001tf}.
Their impact can be removed by implementing kinematical cuts such as
\,$\hat s_{\rm min}=\big(m_D^2-m_\tau^2\big)
\big(m_\tau^2-m_{\mathscr M}^2\big)/m_\tau^2$ \cite{Kamenik:2009kc}.
Our $D_{(s)}^+$ numbers in Eq.\,(\ref{BD2Pff'}) do not yet incorporate them, and we suppose that
they will be taken into account in the experimental searches.

The LQ-induced operators in Eqs.\,\,(\ref{ucff'-R2}), (\ref{ucNN'-tildeR2}), and
(\ref{ucNN'-barS1}) bring about analogous transitions in the charmed-baryon sector.
Here we look at those of the singly charmed baryons $\Lambda_c^+$, $\Xi_c^+$, and $\Xi_c^0$,\,
which have spin parity \,$J^P=1/2^+$,  make up a flavor SU(3) antitriplet, and decay
weakly \cite{Zyla:2020zbs}.
Specifically, we examine the modes \,$\Lambda_c^+\to p\slashed E$\, and
\,$\Xi_c^{+,0}\to\Sigma^{+,0}\slashed E$.\,
The baryonic matrix elements pertaining to the former are \cite{Zyla:2020zbs,Meinel:2017ggx}
\begin{align} \label{Lc2p}
\langle p|\overline u\gamma^\kappa c|\Lambda_c^+\rangle & = \bar u_p^{} \Bigg\{
f_\perp^{} \Bigg[ \gamma^\kappa - \frac{ \textsl{\texttt M}_+^{} \hat p^\kappa
- \textsl{\texttt M}_-^{} \hat q^\kappa }{\hat\sigma_+^{}} \Bigg] + f_+^{}
\Bigg[ \hat p^\kappa - \frac{\textsl{\texttt M}_+^{} \textsl{\texttt M}_-^{}\hat q^\kappa}
{\hat q^2} \Bigg] \frac{\textsl{\texttt M}_+^{}}{\hat\sigma_+^{}}
+ f_0^{} \frac{\textsl{\texttt M}_-^{}\hat q^\kappa}{\hat q^2} \Bigg\} u_{\Lambda_c}^{} \,,
\nonumber \\
\langle p|\overline u\gamma^\kappa\gamma_5^{}c|\Lambda_c^+\rangle & = \bar u_p^{} \Bigg\{
g_\perp^{} \Bigg[ \gamma^\kappa + \frac{ \textsl{\texttt M}_-^{} \hat p^\kappa
- \textsl{\texttt M}_+^{} \hat q^\kappa }{\hat\sigma_-^{}} \Bigg]
- g_+^{} \Bigg[ \hat p^\kappa - \frac{ \textsl{\texttt M}_+^{} \textsl{\texttt M}_-^{}
\hat q^\kappa }{\hat q^2} \Bigg] \frac{\textsl{\texttt M}_-^{}}{\hat\sigma_-^{}} - g_0^{}
\frac{\textsl{\texttt M}_+^{}\hat q^\kappa}{\hat q^2} \Bigg\} \gamma_5^{} u_{\Lambda_c}^{} \,,
\end{align}
where
\begin{align}
\hat p & \,=\, p_{\Lambda_c}^{} + p_p^{} \,, & \hat q & \,=\, p_{\Lambda_c}^{} - p_p^{} \,, &
\textsl{\texttt M}_\pm^{} & \,=\, m_{\Lambda_c}^{} \pm m_p^{} \,, &
\hat\sigma_\pm^{} & \,=\, \textsl{\texttt M}{}_\pm^2 - \hat s \,, ~~~
\end{align}
and $f_{\perp,+,0}$ and $g_{\perp,+,0}$ are form factors depending on $\hat q^2$.
With Eq.\,(\ref{Lc2p}), we derive the amplitude for
\,$\Lambda_c^+\to p\texttt f\bar{\texttt f}{}'$\,
due to ${\cal L}_{uc\texttt{ff}'}$ in Eq.\,(\ref{Lucff'}).
Subsequently, with
\,$|\texttt C_{\texttt{ff}'}^{\textsc v,\textsc a}| =
|\tilde{\textsf c}{}_{\texttt{ff}'}^{\textsc v,\textsc a}| =
G_{\rm F} |\textsc k_{\texttt{ff}'}|/\sqrt8$\,
as before, we arrive at the differential decay rate
\begin{align}
\frac{d\Gamma_{\Lambda_c\to p\texttt f\bar{\texttt f}{}'}^{}}{d\hat s} \,=\,
\frac{\tilde\lambda_{\Lambda_c^{}p}^{1/2}\, G_{\rm F}^2\, |\textsc k_{\texttt{ff}'}|^2}
{768\pi^3m_{\Lambda_c}^3} \Big[ & \hat\sigma_-^{} \Big( f_+^2 \textsl{\texttt M}{}_+^2
+ 2 f_\perp^2 \hat s\Big) + \hat\sigma_+^{}
\Big( g_+^2 \textsl{\texttt M}{}_-^2 + 2 g_\perp^2\hat s \Big) \Big] ~~~ ~~~~
\end{align}
for \,$m_{\texttt f,\texttt f'}\simeq0$,\, in which case the $f_0^{}$ and $g_0^{}$ 
terms drop out from the rate as well.
Its \,$\Xi_c^{+,0}\to\Sigma^{+,0}\texttt f\bar{\texttt f}{}'$ counterparts are similar in form.
Given that the empirical information on \,$\Lambda_c^+\to n\nu e^+$\, and
\,$\Xi_c^{+,0}\to\Sigma^{0,-}\nu e^+$\, is still unavailable \cite{Zyla:2020zbs}, we cannot
implement a procedure like that followed for the meson modes above and must instead rely on
theoretical estimates for the baryonic matrix elements.
Thus, numerically, for the \,$\Lambda_c^+\to p$\, form factors we adopt the results of the lattice
QCD calculation in Ref.\,\cite{Meinel:2017ggx}, while for \,$\Xi_c^{+,0}\to\Sigma^{+,0}$\,
we employ those computed with light-cone QCD sum rules in Ref.\,\cite{Azizi:2011mw}
and assume isospin symmetry.
Putting things together, we then obtain
\begin{align} \label{Lc2pff'}
{\cal B}\big(\Lambda_c^+\to p\texttt f\bar{\texttt f}{}'\big)_{\textsc{lq}}   & \,=\,
2.07 \times 10^{-2}\, |\textsc k_{\texttt{ff}'}|^2 \,, &
\nonumber \\
{\cal B}\big(\Xi_c^+\to\Sigma^+\texttt f\bar{\texttt f}{}'\big)_{\textsc{lq}} & \,=\,
2.39 \times 10^{-2}\, |\textsc k_{\texttt{ff}'}|^2 \,, ~~~ ~~~~
\nonumber \\
{\cal B}\big(\Xi_c^0\to\Sigma^0\texttt f\bar{\texttt f}{}'\big)_{\textsc{lq}} & \,=\,
8.01 \times 10^{-3}\, |\textsc k_{\texttt{ff}'}|^2 \,,
\end{align}
where the $\Lambda_c^+$ and $\Xi_c^{+,0}$ results have uncertainties of order 10\% and 30\%,
respectively \cite{Meinel:2017ggx,Azizi:2011mw}, and the difference between the $\Xi_c^{+,0}$
numbers is ascribable mainly to \,$\tau_{\Xi_c^+}=3.9\,\tau_{\Xi_c^0}$.\,

\end{document}